\documentclass[twocolumn,twoside,slac_two]{revtex4}
\usepackage{graphicx}
\usepackage{fancyhdr}
\usepackage{xspace}
\usepackage{natbib}
\usepackage{color}
\newcommand{\Fermi}{\emph{Fermi}\xspace}
\newcommand{\fermi}{\emph{Fermi}\xspace}
\newcommand{\ltsima} {$\; \buildrel < \over \sim \;$}
\newcommand{\gtsima} {$\; \buildrel > \over \sim \;$}
\newcommand{\lta} {\lower.5ex\hbox{\ltsima}}
\newcommand{\gta} {\lower.5ex\hbox{\gtsima}}
\def\de{$^{\circ}$}

\def\e{\epsilon}
\def\g{\gamma}

\begin{document}

\title{Fermi-LAT Observation of Impulsive Solar Flares}

%

\author{N. Omodei, V. Petrosian}
\affiliation{W. W. Hansen Experimental Physics Laboratory, Kavli Institute for Particle Astrophysics and Cosmology, Department of Physics and SLAC National Accelerator Laboratory, Stanford University, Stanford, CA 94305, USA}
\author{M. Pesce-Rollins}
\affiliation{Istituto Nazionale di Fisica Nucleare, Sezione di Pisa, I-56127 Pisa, Italy}
\author{for the Fermi Large Area Telescope Collaboration}

\begin{abstract}
The Fermi Large Area Telescope (LAT) is the most sensitive instrument ever deployed in space for observing gamma-ray emission $>$100 MeV. This sensitivity has enabled the LAT to detect gamma-ray emission from the Sun during quiescent periods from pions produced by cosmic-ray protons interacting in the solar atmosphere and from cosmic-ray electrons interacting with solar optical photons.  The LAT has detected high-energy gamma-ray emission associated with GOES M-class and X-class X-ray flares accompanied by coronal mass ejections and solar energetic particle events.  In a number of cases, LAT has detected gamma rays with energies up to several hundreds of MeV during the impulsive phase and gamma rays up to GeV energies sustained for several hours after the impulsive phase.  This presentation focuses on observations in the impulsive emission phase in solar flares, including the modest GOES M2-class flare at SOL2010-06-12T0057 and more recent detections, such as the bright X-class flares of March 2012.
\end{abstract}

\maketitle

\thispagestyle{fancy}


\section{Introduction}
The Sun is the most powerful particle accelerator in the solar system and its proximity permits us to investigate over the entire electromagnetic spectrum the processes of particle acceleration and impulsive energy release which occur throughout its cycle. Observations of solar flares over the last few decades have revealed bremmstrahlung, nuclear lines and pion decay components in the spectra of solar flares in the 100 keV to several 10s of MeV energy range. Above 30 MeV the radiation is mainly due to decay of pions however if electrons are accelerated up to 300 MeV, electron bremmstrahlung can also be the source. How the Sun releases this energy, presumably stored in the magnetic fields of the corona,  with such high efficiency, is presently unknown.

Launched in 2008, the \Fermi observatory is comprised of two instruments; the Large Area Telescope (LAT) designed to detect gamma-rays from 20~MeV up to more than 300 MeV~\citep{LATPaper} and the Gamma-ray Burst Monitor (GBM) which is sensitive from $\sim$~8 keV up to 40~MeV \citep{GBMinstrument}. During the first 18 months of operation coinciding with the solar cycle minimum, the \Fermi LAT detected $>$100 MeV gamma-ray emission from the quiescent Sun \citep{abdo11}. As the solar cycle approches it maximum, the LAT has detected several solar flares above 30 MeV during both the impulsive and the temporally extended phases \citep{2011ATel.3635....1O,2011ATel.3552....1O,2012ATel.3886....1T,2012AAS...22042404P,2012AAS...22042403O}.

The first \Fermi GBM and LAT detection of the impulsive GOES M2.0 
flare of 2010 June 12 is presented in~\citet{2012ApJ...745..144A}.
The analysis of this flare was performed using the LAT Low-Energy 
(LLE)~\citet{pela10} 
technique because the soft X-rays emitted during the prompt 
emission of a flare penetrate the anti-coincidence detector (ACD) of the LAT 
causing a pile-up effect which can result in a significant decrease in 
gamma-ray detection efficiency in the standard on-ground photon analysis~\citep{LATPaper}. The pile-up effect has been addressed in detail and we refer the 
reader to \citet{2012ApJ...745..144A} and \citet{LATperform} for a full 
description. The list of all LAT detected flares, and the analysis of the 
first two flares
with long lasting high-energy emission (2011 March 7--8 and 2011 June 7) is 
presented in a paper in preparation. In this paper we present an overview of the \Fermi detection of the GOES M2.0 class flare of 2010 June 12 and the impulsive phase of the GOES X5.4 class flare of 2012 March 7. We also present the localization of the high energy gamma-ray emission of the temporally extended phase of the 2012 March 7 flare.

\begin{figure*}[t]
\centering
\includegraphics[width=\linewidth]{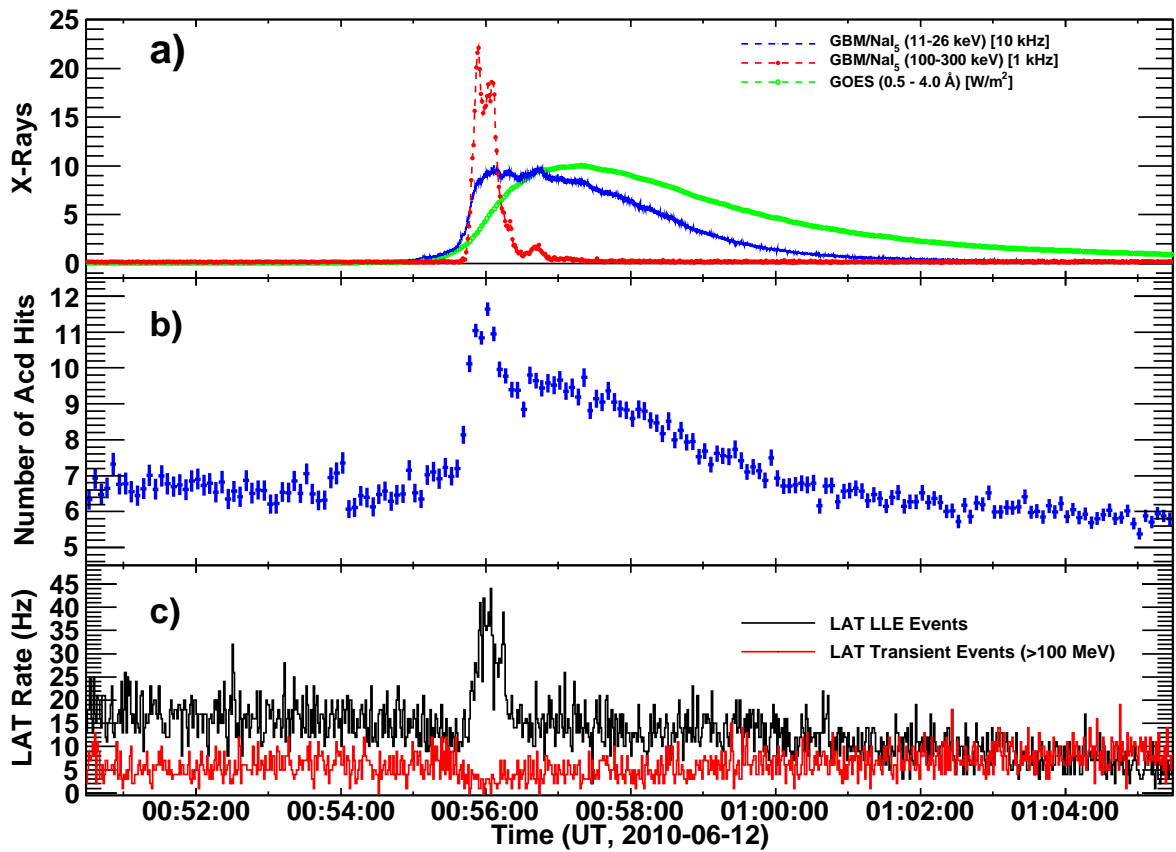}
\caption{Time histories related to the 2010 June 12 solar flare.  a) {\it{GOES}} 0.5 -- 4 $\AA$ rates, and GBM NaI 11 -- 26 keV and 100 -- 300 keV relative rates; b) LAT ACD hit rate $>$100 keV containing contributions from background, $>$100 keV solar flare X rays (impulsive peak) and pulse pile up from 10's of keV solar X rays following the NaI 11 -- 26 keV profile in 1a); and c) LLE and LAT Transient Class event rates. }
\label{lat_nai}
\end{figure*}

\section{The M2.0 flare of 2010 June 12}
On 2010 June 12 00:30 UT a moderate {\it{GOES}} M2.0 class X-ray flare erupted from the active region 11081 located approximately N23$^{\circ}$W43$^{\circ}$. At the time of the flare the \Fermi spacecraft was in sunlight and during a relatively low-background portion of its orbit\footnote{The \Fermi observatory is in a nearly circular orbit with an inclination of 25.6$^{\circ}$ at 565 km.}. The GBM triggered on the flare at 00:55:05.64 UT and detected keV emission for $\approx$10 m. The $11 - 26$ keV emission recorded by the GBM NaI detectors rose precipitously for about 40 s and is shown in Figure \ref{lat_nai}a. For comparison we include the GOES 0.5 -- 4 $\AA$ profile and note that this emission is dominated by 3 keV thermal photons as is reflected in its slower rise and extended tail. The $100 - 300$ keV time profile observed by the GBM's solar facing NaI detector is also plotted in Figure \ref{lat_nai}a. It is clear that the emission peaks more sharply and ends sooner at higher X-ray energies.

The accompanying hard X-ray emission from the flare was detected in the LAT's anti-coincidence detector (ACD) and is reflected in the shape of the average number of ACD tile hits as a funtion of time (shown in Figure \ref{lat_nai}b). The broad peak with a maximum near 00:57 UT of the hit distribution has a shape similar to the 11 -- 26 keV emission observed by the GBM NaI detector and the impulsive peak in the ACD rate is also similar to that observed between 100 and 300 keV by the GBM NaI detector. As is shown in the red curve in Figure \ref{lat_nai}c there is no evidence for the flare in the well-screened standard LAT data products~\citep[shown in the figure are the events belonging to the \texttt{P6TRANSIENT} event class,][]{atwo09}. This is the direct conquence of the pulse-pile effect described in \citep{2012ApJ...745..144A, LATperform}. The black curve in Figure \ref{lat_nai}c is the LAT LLE $>$30 MeV event rate for the time of the flare.

White light emission observed by the Helioseismic and Magnetic Imager (HMI) on the {\it{Solar Dynamics Observatory}} ({\it{SDO}}) \citep{oliv11} in a single 45 s exposure, consistent in time with the hard X-ray emission, revealed two compact footpoints about 10$^4$ km apart.

The $>$30 MeV LLE spectrum of this flare revealed flare emission up to an 
energy of $\sim$400 MeV. The nuclear line emission observed with the GBM 
implies
the presence of accelerated ions up to at least 50 MeV nucleon$^{-1}$.  It is
possible that the flare-accelerated proton spectrum extended up to the
$\sim$300 MeV threshold for pion production.  Alternatively, it is also 
possible
that the LAT emission is from electron bremsstrahlung, either from an
extension to high energies of the electron spectrum producing the X-ray
bremsstrahlung observed in the GBM or from an additional hard electron
component.  One possible way to resolve this ambiguity is to jointly fit the
GBM and LAT spectra assuming different origins for the LAT emission.

\begin{figure}
\centering
\includegraphics[width=\linewidth]{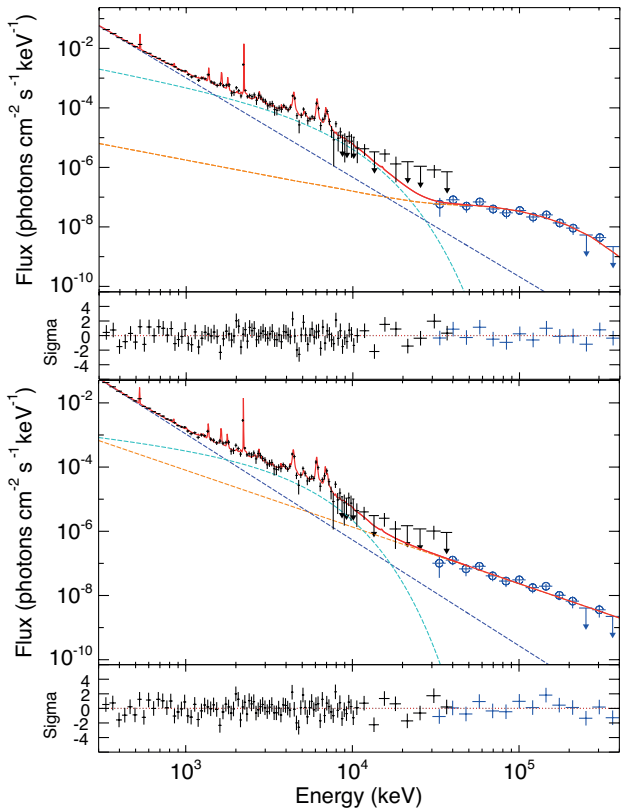}
\caption{
Combined GBM/LAT photon spectrum accumulated between 00:55:40 and 00:56:30
showing the best total fit using the same components as in Figure 3 plus an additional component for the LAT emission.  
The upper panel shows a pion-decay fit to the LAT spectrum; alternatively the lower panel shows a power-law
fit, presumedly representing a third electron bremsstrahlung component.
Note that because this is a photon representation the lines are plotted
at their intrinsic resolution and appear to be more significant than they really are.
}
\label{stack}
\end{figure}
In Figure \ref{stack} we plot the background-subtracted photon spectrum from
0.3 to 400 MeV including both the GBM and LAT data.  We made two fits, using {\em rmfit 3.4}\footnote[1]{R.S. Mallozzi, R.D. Preece, \& M.S. Briggs, ``RMFIT, A Lightcurve and Spectral Analysis Tool'', Robert D. Preece, University of Alabama in Huntsville, (2008): http://fermi.gsfc.nasa.gov/ssc/data/analysis/user/}, customized for the specific solar flare, and the {\em OSPEX}\footnote[2]{SolarSoft: http://www.lmsal.com/solarsoft/} analysis packages, to the joint data sets. In the first fit we assume that the observed LAT emission was from
pion-decay radiation (top panel of Figure \ref{stack}) and the other assuming that it was from a
hard power-law spectrum of electron bremsstrahlung (bottom panel).   Based on
the statistical quality of the fits to the LAT spectrum we cannot distinguish
between the two emission models.  In addition we cannot constrain the origin
of the emission for this event by extrapolating the models into the GBM energy
range; however we note that for a stronger flare we might be able to rule out
a power-law model.  
Also plotted in the figure are the extensions into the LAT energy range of the
power-law and cutoff power-law components derived from the fits to the GBM
data.  The intensities of these components fall at least an order of magnitude
below the LAT measurements and therefore do not make a significant
contribution to the solar emission observed by the LAT.

Even though we cannot statistically distinguish between a pion-decay or electron-bremsstrahlung origin for the observed LAT emission, we can obtain the best-fitting parameters for  these components. 
If the LAT emission is from electron bremsstrahlung, we have found that it cannot be a simple extension of the low-energy bremsstrahlung components that we determined from fits to the GBM data; 
it must be from a distinct population of electrons extending to energies of several hundred MeV. 
However, this high energy electron component would produce a spectrum that steepens beyond tens of MeV due to synchrotron energy losses that increase with energy \citep[see][]{park97}, and must have a quite different origin.
Consequently we believe that this is a less likely scenario than the hadronic model.

Assuming that the LAT emission is from hadronic interactions, we have fit the LAT spectrum with calculated pion-decay templates~\citep{murp87}. The model used to produce these templates depends on the ambient density, composition and magnetic field, on the accelerated-particle composition, pitch angle distribution and energy spectrum. The templates represent a particle population with an isotropic pitch angle distribution and a power-law energy spectrum ($dN/dE \propto E^{-s}$, with $E$ the kinetic energy of the protons) interacting in a thick target with a coronal composition~\citep{ream95} taking  $^4$He/H = 0.1. 
With 67\% confidence (based on $\chi^2$) we conclude that the spectrum of accelerated ions responsible for the pion-decay emission must be steeper than a power-law with index $-$4.5.  We note that there is no change in the quality of the fits for indices steeper than $-5$ due to limited statistics $>$400 MeV. 
We can use the results of our GBM and LAT spectral analyses to obtain information on ions accelerated in the impulsive phase of the June 12 flare.  \citet{murp97} have described how parameters derived from integrated spectroscopic fits and temporal studies can be used to obtain this information.  We first use the nuclear de-excitation line, neutron-capture line, and pion-decay fluences to estimate the overall shape of the accelerated ion spectrum.  
These three emissions are produced by accelerated ions within distinct energy ranges: $\sim$5-20 MeV for the de-excitation lines, $\sim$10-50 MeV for the neutron capture line, and $>$300 MeV for the pion-decay emission.  
Ratios of these emissions therefore determine the relative numbers of accelerated ions in the associated energy ranges.  
We then obtain spectral indices across these energy ranges by comparing measured ratios with ratios from theoretical calculations \citep{murp87,murp05,murp07} based on updated nuclear cross sections.

If we assume that the LAT emission $>$30 MeV was entirely due to pion-decay emission, then we estimate that the flare-accelerated ion spectrum was consistent with a series of power laws, softening with energy, with indices of $\sim$$-3.2$ between $\sim5-50$ MeV, $\sim$$-4.3$ between $\sim$50--300 MeV, and softer than $\sim$$-4.5$ above 300 MeV. 
We summarized in Table~\ref{accelerateddist} our findings, reporting in the first column the emission process responsible for the emission, in the second column the energy range of emitted gamma-rays, in the last two columns we report the energy and the spectral index of the accelerated ions/electrons distributions.

\begin{table}[t]
\begin{center}
\begin{tabular}{lccc}
\hline 
{Component} & {$\gamma$-rays} & {electrons/ions} & Spectral Index \\
& (MeV) & (MeV) & acc. particles\\
\hline 
\hline 
Brem.  & 0.1--1 & 0.1--1 & -3.2 \\
Brem.  & 2--10  & 2--10  & $<$-1.2\\
HE Brem.  & 10--200  & 10--200 & $\approx$-2.0\\
\hline
Neutron Capt.  & 2.2  & 5-50  & $\sim-$3.2\\
Nuclear lines  & 5-20  & 50-300  & $\sim-$4.3\\
Pions & $>$30  & \gtsima 300 & \ltsima$-$4.5\\
\hline
\hline
\end{tabular}
\label{accelerateddist}
\end{center}
\caption{Derived quantities for accelerated particle distributions(with a cut-off at 2.4 MeV)}
\end{table}

\begin{figure*}[t]
\centering
\includegraphics[width=\linewidth]{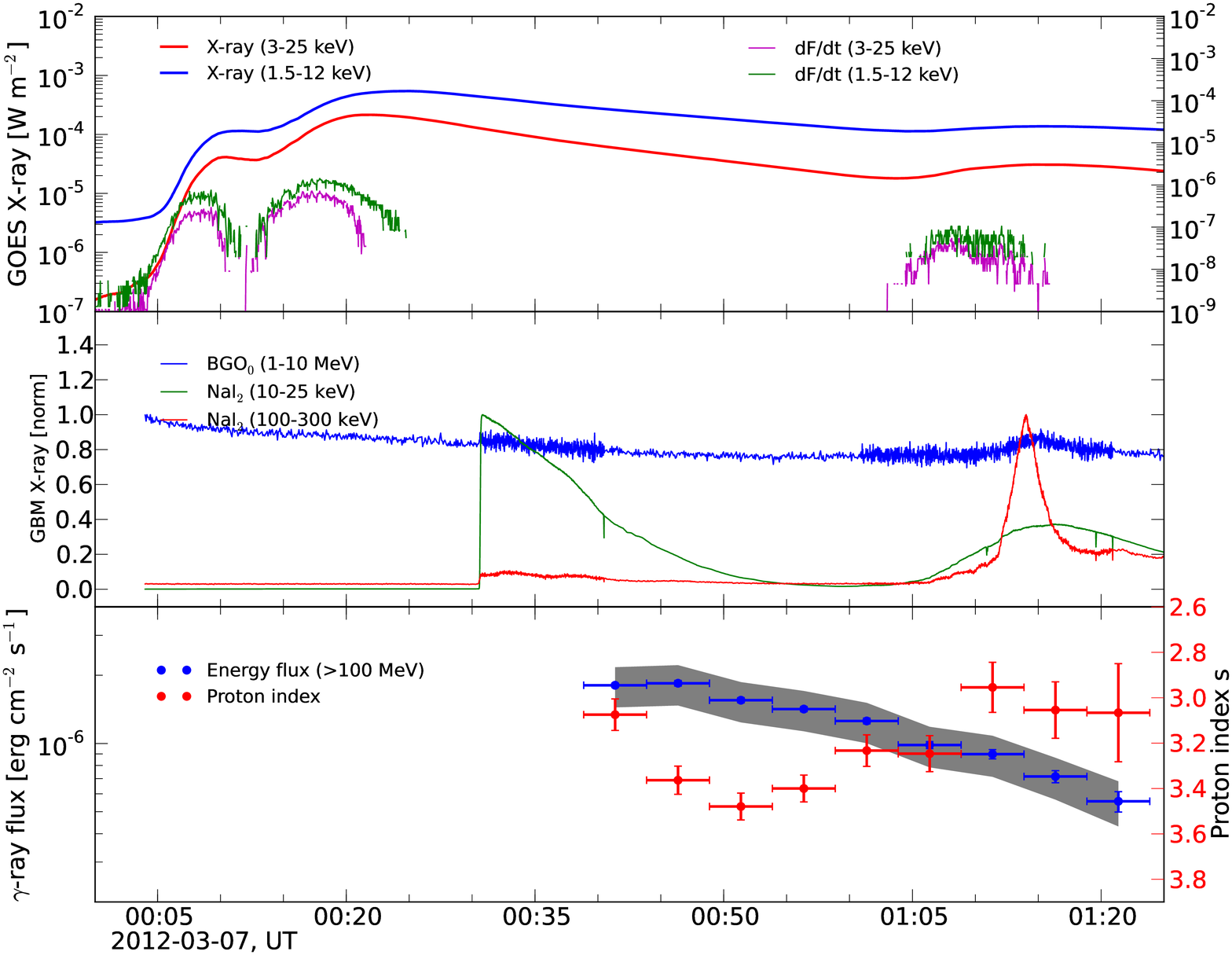}
\caption{PRELIMINARY: Composite light curves for 2012 March 7 flare, covering the first $\sim$80 minutes. 
{\bf Top panel}: Soft X-rays (red:\,1.5--12 keV, blue:\,3--25 keV) from the GOES 15 satellite. On the right axis are the first derivatives of the soft X-rays fluxes (magenta:\,1.5--12 keV, green:\,3--25 keV).
These curves approximate accelerated electron impulsive lightcurves \citep{1968ApJ...153L..59N}.
{\bf Middle panel}: Hard X-rays count rates from the GBM; green and red for NaI$_{2}$ 10--25\,keV and 100--300\,keV energy channels, and blue for the BGO$_{0}$ detector. 
{\bf Bottom panel}: LAT ($>$100 MeV) gamma-ray flux (blue) and derived proton spectral index (red). 
The gray band represents the systematic uncertainties associated to the flux measurement, and it is obtained by adding 20\% systematic error in quadrature.}
\label{LC20130307}
\end{figure*}

\section{The X class flares of 2012 March 7}

On 2012 March 7 two bright X-class flares originating from the active region NOAA AR\#:11429 (located at N16$^{\circ}$E30$^{\circ}$) erupted within an hour of each other, marking one of the most active days of Solar Cycle 24. 
The first flare started at 00:02:00 UT and reached its maximum intensity (X5.4) at 00:24:00 UT while the second X1.3 class flare occurred at 01:05:00 UT, reaching its maximum 9 minutes later. 

The GOES satellite observed intense X-ray emission beginning at about 00:05:00 UT and lasting for several hours. Moreover, it detected in three energy bands Solar Energetic Particle (SEP) protons originating from the same flares. 
In the top panels of Figure~\ref{LC20130307} the X-ray data from GOES 15 satellite measured in both 3--25 keV and 1.5--12 keV  channels are shown for two time intervals during the flaring episode. GOES soft X-ray light curves usually do not follow the impulsive nature of the activity because they trace the accumulated energy input by the accelerated particles. In general, based on the so-called \emph{Neupert effect} \citep{1968ApJ...153L..59N} the derivative of these light curves is considered to be a good proxy for the temporal evolution of the spectrum of accelerated particles. Figure~\ref{LC20130307} shows the light curves for the 1.5--12 and 3--25 keV GOES bands, together with their corresponding derivatives. Such derivatives make it clear that the first flare consisted of two impulsive bursts with a duration of a few minutes each while the second flare was composed of only one such pulse.
Unfortunately the Reuven Ramaty High-Energy Solar Spectroscopic Imager \citep[RHESSI,][]{2002SoPh..210....3L} was not observing the Sun during this period.

Orbital sunrise of \Fermi occurred less than six minutes after the peak of the first flare, triggering the GBM at 00:30:32.129 UT  (coinciding with the abrupt rate increase visible in the middle panel of Figure~\ref{LC20130307}).
The second flare is also clearly visible in the BGO$_{0}$ detector of the GBM\footnote{We use dead-time corrected count rates from the NaI$_{2}$ in the 10--25\,keV and 100--300\,keV energy bands and from the BGO$_{0}$ in the 1--10\,MeV energy band.}.
The \Fermi LAT $>$100 MeV count rate was dominated by the gamma-ray emission from the Sun\footnote{http://apod.nasa.gov/apod/ap120315.html}, which was nearly 100 times brighter than the Vela Pulsar in the same energy range.
During the impulsive phase (the first eighty minutes) the X5.4 flare was so intense that the LAT observation suffered from the pile-up effect \citep[see][]{2012ApJ...745..144A},
so the standard instrument response functions (IRFs) could not be used. Instead, we used the LLE technique to analyze the impulsive phase of this bright flare.

We fit the data using \texttt{XSPEC}\footnote{http://heasarc.gsfc.nasa.gov/docs/xanadu/xspec/index.html} to test three models. The first two are simple phenomenological functions, to describe bremsstrahlung emission from accelerated electrons, namely a pure power law (PL) and a power law with an exponential cut-off (EXP):
\begin{equation}
 \frac{dN(E)}{dE} = N_{0}\,\e^{-\Gamma}\,\exp\left({-\frac{E}{E_{co}}}\right);
\label{eq1}
\end{equation} 
where $\Gamma$ is the photon index and $E_{co}$ is the cut-off energy.
We found that the data clearly diverge from a pure power law spectrum and that the EXP provides a better fit in all time intervals considered.
The third model uses the same templates used for the 2010 June 12 flare, which are based on a detailed study of the gamma rays produced from pion decay~\citep{murp87}.
When using the pion-decay templates to obtain the gamma-ray flux value we fit the data varying the proton spectral index $s$ from 2--6, in steps of 0.1. 
In this way, we fit the LAT data with a model with two free parameters, the normalization and the proton index $s$.
The time dependence of the $>$100\,MeV gamma-ray flux and of the proton index $s$, derived using gamma-ray LAT data, is displayed in the lower panel of Figure~\ref{LC20130307}.
It appears that, after a short phase of spectral softening, the proton spectral index hardens before the start of the impulsive phase of the second flare as seen by the GBM detectors (middle panel of Figure~\ref{LC20130307}). 
The spectral index $s$ correlates better with the GBM flux than with the flux at high energy measured by the LAT.

\subsection{Localizing the high-energy gamma-rays}
\begin{figure}[t]
\centering
\includegraphics[width=\linewidth]{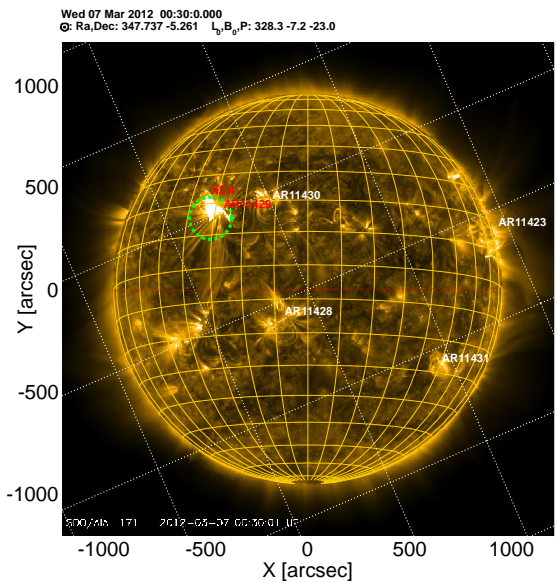}
\caption{PRELIMINARY: Location of the gamma ray emission above 100 MeV for the time-integrated analysis. 
The image on the background are from SDO (AIA 171\AA) and are taken at the time of the flaring episode. 
Active regions are flagged with their respective NOAA numbers. The region marked in red is the region associated with the X-class flares of 2012 March 7, located at N16E30 (X,Y=-471,373 arcsec).
The green circle is the 68\% error radius (+systematic error added in quadrature). The grid on the background is the coordinate grid of equatorial coordinates, while the yellow sphere is the heliocentric coordinate grid (with the projected Sun rotation axis parallel to the Y-axis, the Z-axis is the line of sight (from the Sun to the observer) and the X-axis in the cartesian projection complete the normal base.}
\label{loc20130307}
\end{figure}

High energy gamma-ray emission measured by the LAT during the 2012 March 7 events lasted several hours (approximately 20), providing very strong constrain on the nature of the emission processed and on the origin of the accelerated particles.
Since the effect of the pulse pile-up after the eighty minutes of LAT observations is negligible, we can adopt standard likelihood analysis on the time extended emission to perform spectral analysis and also to estimate the localization of the high-energy gamma-rays.
Figure~\ref{loc20130307} shows the localization for the entire period of the high energy emission not affected by ACD pulse pile-up.
The uncertainties on the localization are obtained by combining the 68\% error radius from \texttt{gtfindsrc} with the systematic error associated with the ``fisheye'' effect in quadrature \citep{2012ApJS..203....4A}. 
We estimate the latter using Monte Carlo simulations and find it to be 0\de.02.
The reconstructed location of the emission is consistent with the direction of the Sun. A more detailed time-resolved analysis will be presented in a forthcoming paper.

\section{Conclusions}
The sensitivity of the \Fermi-LAT enables the investigation of several aspects of solar flares that were not previously accessible.
In this conference proceedings we have anticipated some results focusing on the impulsive phase, although a more comprehensive presentation will be the subject of a forthcoming paper.
Results on the 2010 June 12 flare \citep{2012ApJ...745..144A} have already shown some of the potentialities for solar flare observations by the \Fermi telescope, in particular the capability to observe gamma-ray broad band spectrum, 
and to derive parameters for the underlying distribution of accelerated particles. 
The data for the exceptionally bright solar flares of 2012 March 7 represent an excellent opportunity to study the details of these characteristics.
For this flare we were able to constrain the model for the emission at high-energy using \Fermi-LAT data, deriving the temporal evolution of the underlying population of accelerated protons/ions producing $\g$-rays via pion decay.
For the first time we were able to localize the $\g$-ray emission, which is consistent with the region of the hard X-ray flare, suggesting that $\g$-ray emission is produced by accelerated protons/ions interacting with the Sun chromosphere and below. More details on the analysis and different scenarios for particle acceleration in solar flares will be extensively covered in a subsequent publication.

\bigskip 
\begin{acknowledgments}
The \fermi~LAT Collaboration acknowledges generous ongoing support from a number of agencies and institutes that have supported both the development and the operation of the LAT as well as scientific data analysis. These include the National Aeronautics and Space Administration and the Department of Energy in the United States, the Commissariat \`a l'Energie Atomique and the Centre National de la Recherche Scientifique / Institut National de Physique Nucl\'eaire et de Physique des Particules in France, the Agenzia Spaziale Italiana and the Istituto Nazionale di Fisica Nucleare in Italy, the Ministry of Education, Culture, Sports, Science and Technology (MEXT), High Energy Accelerator Research Organization (KEK) and Japan Aerospace Exploration Agency (JAXA) in Japan, and the K.~A.~Wallenberg Foundation, the Swedish Research Council and the Swedish National Space Board in Sweden. \\
Additional support for science analysis during the operations phase is gratefully acknowledged from the Istituto Nazionale di Astrofisica in Italy and the Centre National d'\'Etudes Spatiales in France.

\end{acknowledgments}

\bigskip 
\bibliography{Fermi2012}
\end{document}